\newcommand{\f}{\mbox{$ \varphi $}}

\newcommand{\p}{\mbox{$ \partial $}}

\newcommand{\Beq}{\begin {equation}}
\newcommand{\Endeq}{\end {equation}}

\newcommand{\Beginc}{\begin {center}}
\newcommand{\Endc}{\end {center}}

\newcommand{\C}{\cite}

\documentstyle[12pt]{article}
\font\titlefont=cmbx10 scaled \magstep3

\begin{document}

\begin{flushright}
\vspace*{-2cm}
TUTP-93-  \\ Dec. 1993
\vspace*{2cm}
\end{flushright}

\begin{center}
{\titlefont EVOLUTION OF TOPOLOGICAL DEFECTS DURING INFLATION}\\
\vskip .7in
Rama Basu  \& Alexander Vilenkin\\
\vskip .3in
Institute of Cosmology\\
Department of Physics and Astronomy\\
Tufts University\\
Medford, Massachusetts 02155\\
\vskip 30pt
\end{center}

\vskip 1in
\newpage
\begin{abstract}
\baselineskip=24pt

Topological defects can be formed during inflation by phase transitions
as well as by quantum nucleation.
We study the effect of the expansion of the Universe
on the internal structure of the defects.
We look for stationary solutions to
the field equations, i.e. solutions that depend only on the proper
distance from the defect core. In the case of very thin defects,
whose core dimensions are much smaller than the de Sitter horizon,
we find that the solutions are well approximated by the flat space solutions.
However, as the flat space thickness
parameter $\delta_0$ increases we notice a deviation from this, an effect that
becomes dramatic
as $\delta_0$ approaches $(H)^{-1}/{\sqrt 2}$. Beyond this critical value
we find no stationary solutions
to the field equations. We conclude that only defects that have flat space
thicknesses less than the critical value survive, while thicker defects are
smeared out by the expansion.

\end{abstract}

\newpage
\baselineskip=24pt
\Beginc
{\bf I. Introduction}
\Endc

The inflationary epoch in cosmology increases the size of the universe by
a huge factor \C{I}, implying that the present observable universe originated
from
a tiny initial region. Any topological defects formed at the onset of
inflation, or before the start of inflation would be inflated away, implying
that the only defects that could possibly be observable at the present time
would be those formed at or near the
end of the inflationary epoch.
Topological defects can be
continuously formed during the course of inflation by quantum mechanical
tunnelling processes\C{bgv}, and defects formed during
inflation by this mechanism could be present after inflation
with appreciable densities. Phase transitions could also occur during
inflation if the symmetry breaking field \f~ is coupled to the inflaton
field \C{VS}. The characteristic length scales of the defects formed in such
phase transitions increases exponentially due to inflation.
However, if the phase transition
occurs close enough to the end of inflation, so that this length scale does
not exceed the size of the presently observable universe, then these defects
are not diluted away. Defects could also be formed during inflation
by quantum fluctuations \C{HP} in the
case where the symmetry is broken before or at
the beginning of inflation.

All of these defects would have been exposed to the exponential expansion of
the
Universe long enough to make the question
of what happens to the internal structure of these defects during inflation
significant. In this paper we will address this question.
The inflationary universe is approximated by de Sitter spacetime, which
has a constant expansion rate $H$. We can therefore
look for stationary solutions to the
scalar field equations for domain walls, strings and monopoles.

Intuitively, one expects that the defect structure in de Sitter
space will be essentially the same as in flat space if the flat
space thickness of the defects, $\delta_0$, is much smaller than
the de Sitter horizon, $H^{-1}$. On the other hand, it is hard to see
how a coherent defect structure can be sustained on scales greater than
$H^{-1}$, and we expect that stationary solutions do not exist when $\delta_0$
exceeds some critical value $\delta_{\rm c}\sim H^{-1}$, and that for
$\delta_0 > \delta_{\rm c}$, the defects are smeared by the expansion
of the universe. We shall see that these expectations are indeed correct.

This paper is organized as follows :
We study the structure of defects using a simple
scalar field model; Sec II deals with the structure of
domain walls, for which we obtain numerical solutions. We also obtain
analytic solutions in two asymptotic regimes
corresponding respectively to the case of
very thick walls with flat space wall thickness comparable to the critical
value and to the case of very thin walls. We do a similar analysis for
strings and monopoles in Sec. III. Our conclusions are
summarized in Sec. IV.

\Beginc
{\bf II.  Domain Walls }
\Endc

The spacetime of the inflationary universe is approximately de Sitter,
and the
metric with flat spatial sections is given by
\begin {equation}
ds^2 = dt^2 - e^{2Ht}[ dx^2 + dy^2 + dz^2 ]
\end {equation}
We consider a one component scalar field theory with a simple double well
potential
\Beq
V(\f) = {\lambda \over 2}( \f^2 - \eta^2 )^2
\Endeq

The scalar field equation is given by
\Beq
{1\over {\sqrt {-g}}}\p_{\mu}(g^{\mu \nu }\p_{\nu}\f )
=  -2 \lambda \f ({\f }^2 - \eta^2)
\label{eq:scalarfldeqone}
\Endeq

We consider a plane domain wall situated at $z=0$. This is not as
special a case as it appears, because a recoordinatization of deSitter
space in the calculations of \C{bgv}, reveals \C{jv} that the spherical
domain walls nucleating during inflation are equivalent to plane domain
walls appearing at $ z=0$ in the new coordinates.
We can write out the field equation
explicitly, as follows.
\Beq
{\p^2 \f \over \p t^2} + 3H{\p \f \over \p t } - \exp {-2Ht} \{
{\p^2 \over \p x^2} + {\p^2 \over \p y^2}  + {\p^2 \over \p z^2} \} \f
=  -2 \lambda \f ({\f }^2 - \eta^2)
\Endeq

If the wall is not smeared by the expansion, it should be described by
a stationary field configuration, in terms of the proper distance from
the wall.
Accordingly, we choose the following ansatz for
\f :

\Beq
\f = \f (u), \quad {\rm where } \quad u = Hz\exp {Ht}
\label{eq:wallansatz}
\Endeq

In terms of the variable $u$, the field equation may be rewritten
as

\Beq
(1-u^2) {\p^2 y \over \p u^2} - 4u {\p y \over \p u} = 2C y( y^2 -1)
\label {eq:fieldeq1}
\Endeq

where $y = \f /\eta $ and $C = H^{-2}/ \lambda^{-1}\eta ^{-2} = H^{-2}/
\delta_0^2$ where $\delta_0 =({\sqrt \lambda }\eta )^{-1} $
is the flat-space wall thickness. The solution
must obey the boundary conditions
\Beq
y(0) = 0 \quad y(\pm \infty) = \pm 1.
\label{eq:bcw}
\Endeq

Eq. (\ref{eq:fieldeq1}) cannot be solved in closed form. We therefore
used a shooting routine to obtain numerical solutions. The numerical results
are graphically depicted in Fig. 1. As expected, at large $u$ the field
$\f$ approaches its VEV $\eta $. However, the solutions exhibit an
aperiodical damped oscillatory behaviour as \f~ approaches the VEV,
as opposed to the monotonic approach to the VEV in flat space.
At large values of $C$, when $\delta_0 \ll H^{-1}$, the solution
is essentially identical to the flat space solution,
$y = {\rm tanh} ({\sqrt C} u)$. (In this case most of the variation
of $y$ between 0 and 1 occurs at small values of $u$, so that the
terms proportional to $u^2$ and $u$ in Eq. (\ref{eq:fieldeq1}) are
negligible, and it reduces to the corresponding flat-space equation).
As $C \longrightarrow 2$, or $\delta_0 \longrightarrow H^{-1}/\sqrt 2$,
the solution approaches $y=0$ over the whole range of integration.
We found no non-trivial solutions to
the field equations when $\delta_0 \ge H^{-1}/\sqrt 2$.
We now wish to examine the asymptotic behaviour of the solutions more closely.
Accordingly, we consider two asymptotic regimes, one where $y \ll 1$
corresponding to very thick walls with flat space wall
thickness $\delta_0 \approx H^{-1}/\sqrt 2$ and
the other where $|y-1|\ll 1$, which corresponds to
large distances  from the core.
\vskip 4pt

\Beginc
 {a) Large-distance asymptotic}
\Endc

In this asymptotic region where $u $ is large, the field equation
can be written as

\Beq
u^2 {\p^2 y \over \p u^2} + 4u {\p y \over \p u} +  2C y( y^2 -1) = 0
\label{eq:asym1}
\Endeq

 Furthermore, we can
assume that $y = 1 - f $ where $f \ll 1$, in the region of interest.
Substituting this in Eqn. (\ref{eq:asym1}) and discarding higher order terms
in $f$ we have

\Beq
u^2 {\p^2 f \over \p u^2} + 4u {\p f \over \p u} + 4C f = 0
\label{eq:domasym}
\Endeq

Now, making a change of variables $v=\ln u$, we have

\Beq
{\p^2 f \over \p v^2} + 3 {\p f \over \p v} + 4C f = 0
\label{eq:domv}
\Endeq

The solution is of the form $\exp ({\alpha v})$, with
\Beq
\alpha^2 + 3\alpha + 4C = 0
\Endeq
It follows that
\Beq
\alpha = -{3 \over 2} \pm {{\sqrt {9 -16C}}\over 2}
\Endeq

The solution for $f$ is therefore given by,
\Beq
f = A u^{-{3\over 2}}\cos \Bigl ({{1 \over 2} {\sqrt {16C- 9}}\ln u}\Bigr )
\label{eq:thinwall}
\Endeq

This indicates that the field $\f $ approaches the vacuum expectation value
$\eta $ for large $u$ as expected. In agreement with our numerical results,
the field exhibits a
damped oscillatory behaviour
about the vacuum expectation value for large $u$.

\Beginc
b) Near-critical behavior ($\delta_0 \approx H^{-1}/\sqrt 2$)
\Endc

The numerical results presented earlier in this section indicate
that $y = \f /\eta $ becomes very small over the whole range of
integration of Eq. (\ref{eq:fieldeq1}) when $C\approx 2$, i. e.
when $\delta_0 \approx H^{-1}/\sqrt 2$.
In this regime, we are therefore justified to assume that $y \ll 1 $ upto
large values of $u$. The
field equation can then be linearized to become,

\Beq
(1-u^2) {\p ^2 y \over \p u^2} - 4u{\p y \over \p u} + 2C y = 0
\label{eq:ysmall}
\Endeq

Now making a change of dependent variable $ y = {w \over \sqrt { (1- u^2)}}$
and replacing in Eq. (\ref{eq:ysmall}) we have

\Beq
(1-u^2) {\p ^2 w \over \p u^2} - 2u{\p w \over \p u} + \{ 2(C +1)- (1-
u^2)^{-1}
\}w = 0
\label{eq:legendre}
\Endeq

This is precisely the associated Legendre equation \C{AS}. The general
solution to this equation is

\Beq
w(u) = A {P^\mu}_\nu  + B {Q^\mu}_\nu
\Endeq
with $\mu = 1$ and $\nu ( \nu + 1) = 2 ( C+1) $.

In order that the solutions to this equation be bounded at $u=\pm 1$ ,
$\nu $ is constrained to assume only integral values. The value of
interest to us is $\nu = 2 $ which corresponds to
$C=2$. With this value, the non-singular solution of Eq. (\ref{eq:legendre})
satisfying the boundary condition (\ref{eq:bcw}) at $u=0$ is
\Beq
y= A(1-u^2)^{-1/2}P^1_2(u)=Au
\label{eq:thicksoln}
\Endeq
where $A$ is a constant. Although (\ref{eq:thicksoln}) solves
Eq. (\ref{eq:legendre}) only for $C=2$, we expect it to be
approximately valid for $C \approx 2$.

The constant $A$ was approximately
evaluated in our earlier paper \C{bv} as
\Beq
A^2={7\over 3C}(C-2)
\label{eq:constant}
\Endeq
In that paper we studied instanton solutions of the scalar field equations
in Euclideanized de Sitter space. These instantons describe nucleation
of spherical domain walls during inflation. The subsequent evolution of the
walls can be found by analytically continuing the instanton solutions.
Moreover, by  a suitable choice of coordinates, an expanding nucleated wall
can be transformed into a planar wall (\ref{eq:wallansatz}). [This
transformation is similar to the transformation from the de Sitter metric with
closed spatial sections to the spatially flat form.] This leads to the
conclusion that the expression (\ref{eq:constant}) for A is still applicable
for Lorentzian wall solutions.

The wall thickness in de Sitter space can be approximately calculated
using the relation
\Beq
y'(0)\delta \sim 1
\label{eq:thickness}
\Endeq

Combining Eqs. (\ref{eq:thicksoln}), (\ref{eq:constant}) and
(\ref{eq:thickness}) we can obtain an
approximate expression for the wall thickness in de Sitter space

\Beq
 \delta \sim H^{-1} \{1 - 2 (H\delta_0 )^2 \}^{-{1 \over 2}}
\label{eq:wallthick}
\Endeq

The effect of de Sitter expansion assumes significance as
$\delta_0 \sim H^{-1}$,
and becomes dramatic when $\delta_0 \longrightarrow { H^{-1}\over {\sqrt 2}}$.
In this limit Eq. (\ref{eq:wallthick}) indicates that the wall thickness
grows without bound. Thicker walls cannot survive in de Sitter space as
coherent objects. They are smeared by the expansion of the universe.

\Beginc
{\bf III.  Strings and Monopoles  }
\Endc

\Beginc
 a) Strings
\Endc

We again consider a scalar field theory with a simple double well potential
\Beq
V(\f_a) = {\lambda \over 2}( \f_a \f_a - \eta^2 )^2
\Endeq
where $a=1,\ldots ,N$. The values $N=2$ and $N=3$ correspond to
strings and monopoles respectively.
The scalar field equation is given by
\Beq
{1\over {\sqrt {-g}}}\p_{\mu}(g^{\mu \nu }\p_{\nu}\f_a )
=  -2 \lambda \f_a ({\f_b}\f_b - \eta^2)
\label{eq:scalarfldeq}
\Endeq

We  consider an infinite straight string situated along the $z$-axis.This
is again a configuration that is equivalent to the nucleating circular
loops discussed in \C{bgv}. The cylindrical symmetry of this string
configuration suggests a recoordinatization of the de Sitter metric as follows:

\Beq
ds^2 = dt^2 - \exp{(2Ht)}(d\rho^2 + dz^2 + \rho^2 d\phi^2 )
\label{eq:metriccy}
\Endeq

The string is described by a two-component scalar field theory.
Since we are looking for stationary solutions to the scalar field
equation, we choose the following ansatz for the scalar field :

\begin{eqnarray}
\f_1 & = &f(\rho e^{Ht})\cos (n \phi) \nonumber \\
\f_2 & = &f(\rho e^{Ht})\sin (n \phi)   \label{eq:stringansatz}
\end{eqnarray}

Replacing $ \f_a $ from Eq. (\ref{eq:stringansatz}) in
Eq. (\ref{eq:scalarfldeq}), and introducing the dimensionless variables
$u = H\rho e^{Ht} $ and $y = f/\eta$, the two field equations reduce to
a single equation for $y$ :

\Beq
(1 - u^2) {\p ^2 y \over \p u^2 } + { (1 - 4u^2) \over u} {\p y \over \p u}
- y u^{-2} = 2Cy(y^2-1)  \label{eq:stringeq}
\Endeq

where $C = {H^{-2} \over \delta_0^2}$, and $\delta_0$ is the flat space
thickness of the string core. As we did for walls,
we look for solutions to Eq. (\ref{eq:stringeq})
in two asymptotic regimes, i.e.
the region far away from the string core where $y \approx 1$, and thick string
asymptotics, where $y\approx 0 $.

In the large distance regime, $u \gg 1$, and $y = 1+f$, with $f\ll 1$.Then
Eq. (\ref{eq:stringeq}) reduces to the same equation (\ref{eq:domasym})
that we obtained in the case of a domain wall. The solution is given by Eq.
(\ref{eq:thinwall}). Once again, instead of a monotonic approach to $\eta$
as observed in flat space, the string solution exhibits a damped oscillatory
behavior.

For a thick string we can assume that $y\ll 1$ and linearize Eq.
(\ref{eq:stringeq}) by discarding the cubic term in $y$. We expect that in this
case, the solution will be well approximated by a linear term,
\Beq
y = Au
\label{eq:linearu}
\Endeq
up to very large values of $u$. Substituting this in the linearized
equation Eq. (\ref{eq:stringeq}), we obtain a condition for $C$, $C=2$.
This indicates that the critical value of the flat-space core thickness
is $\delta_0 = H^{-1}/\sqrt 2$, the same as for the domain wall.

\Beginc
b)  Monopoles
\Endc

We consider a monopole located at the origin.
Using spherical polar coordinates to
describe the spatial part of the de Sitter metric,

\Beq
ds^2 = dt^2 - \exp{(2Ht)}(dr^2 + r^2 d\theta^2 + r^2 \sin^2 \theta d\phi^2 )
\label{eq:metricsp}
\Endeq

we choose the following ansatz for the three components of the scalar field :

\begin{eqnarray}
\f_1 &=& f(r e^{Ht})\cos \theta \nonumber \\
\f_2 &=& f(r e^{Ht})\sin \theta \cos \phi \nonumber \\
\f_3 &=& f(r e^{Ht})\sin \theta \sin \phi  \label{eq:monoansatz}
\end{eqnarray}

The field
equation then reduces to

\Beq
(1 - u^2) {\p ^2 y \over \p u^2 } + { (2 - 4u^2) \over u} {\p y \over \p u}
- 2y u^{-2} = 2Cy(y^2-1)  \label{eq:monoeq}
\Endeq

where $u = Hre^{Ht}$, $y = f/\eta$ and $C$ has the same meaning as before.
In the large distance limit, we find once again that the asymptotic behavior
is described by Eq. (\ref{eq:domasym}), with the solution (\ref{eq:thinwall}).

In the thick monopole limit, linearizing Eq. (\ref{eq:monoeq}), and
substituting the linear $u$-dependence (\ref{eq:linearu}), we obtain $C=2$
as in the previous two cases.

Instead of using the linear ansatz (\ref{eq:linearu}) for thick monopole and
string solutions, we could look for a general solution of the linearized field
equation and require regularity at the horizon $(u=1)$, as we did for domain
walls in Section II. The analysis is essentially identical to that given in
Ref. \C{bv} and leads to the same critical value of
$C=2$.%%%%%%%%%%%%%%%%%%%%%%%%%%%%%%%%%%%%%%%%%%%%%%%%%%%%%%%%%%%%%%%%%%%%%%%
%%%%%%%%%%%%%%%%%%%%%%%%%%%%%%%%%%%%%%%%%%%%%%%%%%%%%%%%%%%%%%%%%%%%%%%
\vskip 15pt

\Beginc
{\bf IV.  Summary  }
\Endc

We studied the effect of the exponential expansion of the Universe
on the internal structure of topological defects, concentrating
mainly on the case of domain walls. We found that flat-space domain wall
solutions whose thickness, $\delta_0$, is much smaller than the de Sitter
horizon, $H^{-1}$, are not substantially modified when the wall is transplanted
to de Sitter space. the main modification is that at distances greater than
$H^{-1}$ from the wall, the field exhibits a damped
oscillatory approach to the VEV $\eta$ (in contrast with the monotonic
approach to $\eta$ in flat space). As the flat-space wall thickness
$\delta_0$ approaches the critical value
\Beq
\delta_{\rm c}= H^{-1}/\sqrt 2
\label{eq:critical}
\Endeq
the effect of de Sitter
expansion becomes predominant. In this limit, the wall thickness grows
unboundedly according to the relation (\ref{eq:wallthick}). No regular
solutions to Eq. (\ref{eq:scalarfldeqone}) exist for
$\delta_0 > \delta_{\rm c} $.
Very similar results have been obtained for strings and monopoles. In all three
cases, when the flatspace thickness of the defect core $\delta_0$ exceeds the
critical value (\ref{eq:critical}), no stationary solutions
exist indicating that thicker defects cannot survive in de Sitter space as
coherent objects. They are smeared out by the expansion of the universe,
and their thickness grows as $\delta \sim e^{Ht}$.

The formation of defects by quantum fluctuations during inflation, which
was discussed in Refs. \C{HP}, assumes that the scalar field mass
$m_{\f} \ll H$, so that $\delta_0 \sim m_{\f }^{-1}\gg H^{-1}$.
Hence, in this case, defects cannot be considered as 'formed' until
after the end of inflation, and instead of defects it is more appropriate
to describe them as `zeros of the field \f'. However, since defect formation
eventually occurs where there are zeros of \f, the conclusions of Refs. \C{HP}
remain essentially unchanged.

\newpage

\Beginc
Figure Caption
\Endc

\vspace*{1in}

Fig. 1. The scalar field $\f/\eta$ as a function of $zH\exp (Ht)$, shown
for different values of the flat-space thickness parameter, $C=10$, $C=4$,
$C=2.5$, $C=2.05$, and $C=2.001$


\begin{thebibliography}{--}
\bibitem{I}For a review of inflation, see,
for e.g., E.W. Kolb and M.S. Turner, {\it The
Early Universe} (Addison-Wesley, New York,1990); A. Linde,
{\it Particle Physics and Inflationary Cosmology}
(Harwood Academic, Chur, Switzerland, 1990);
or S. K. Blau and A. H. Guth, in {\it 300 Years of Gravitation},
ed. by S. W. Hawking, and W. Israel (Cambridge University Press, Cambridge,
England, 1987).

\bibitem{bgv}R. Basu, A. H. Guth and A. Vilenkin, Phys. Rev. {\bf D44},
340 (1991).

\bibitem{VS}Shafi and Vilenkin, Phys. Rev. {\bf D29}, 1870 (1984);
L. Kofman and A. D. Linde, Nucl. Phys. {\bf B 282}, 555 (1987).

\bibitem{HP}E. T. Vishniac, K. A. Olive and D. Seckel, Nucl. Phys. {\bf B289},
717 (1987),
A. D. Linde and D. H. Lyth, Phys. Lett. {\bf B246}, 353 (1990),
H. M. Hodges and J. R. Primack, Phys. Rev. {\bf D43}, 3155(1991),
M. Nagasawa and J. Yokoyama, Nucl. Phys. {\bf B370}, 472 (1992).

\bibitem{jv}J. Garriga and A. Vilenkin, Phys. Rev. {\bf D45}, 3469 (1992).

\bibitem{AS}M. Abramowitz and I. A. Stegun, {\it Handbook of
Mathematical Functions} (Dover, New York, 1965).

\bibitem{bv}R. Basu and A. Vilenkin, Phys. Rev. {\bf D46}, 2345 (1992).

\end{thebibliography}
\end{document}